# A Model for Thermodynamic Properties of Monoatomic Liquids

Drew Lilley[1,2], Anubhav Jain[1,*], Ravi Prasher[1,2,*]

[1]Energy Technologies Area, Lawrence Berkeley National Laboratory, Berkeley, CA 94720, USA
[2]Department of Mechanical Engineering, University of California, Berkeley, CA 94720, USA
* e-mail: ajain@lbl.gov, rsprasher@lbl.gov

We present an analytical model for calculating the thermodynamic properties of monoatomic liquids using a rough potential energy surface (PES). The PES is transformed into an equivalent simple harmonic oscillator. Without employing any adjustable parameters, the model agrees closely with experimental entropy, heat capacity, and latent heat of fusion/vaporization data for monatomic liquids. In addition, it offers a simple, physical explanation for Richard's Melting rule, and provides a material-dependent correction to Trouton's Vaporization rule.

A general approach to calculating the thermodynamic properties of liquids has been a long-standing problem in condensed matter physics [1]. Its solution would offer scientists and engineers insight into reaction kinetics, macroscopic thermodynamic cycles, and material design. It is also needed for the prediction of first- and second-order phase transitions involving liquids. For example, there is currently not even a theory for calculating the entropy of fusion without fitting or adjustable parameters [2]. Such a theory is needed to guide the discovery and development of phase change-based thermal storage materials among other applications.

There is a long history of insightful contributions into the nature and theory of the liquid state. Borne and Green [3] constructed a general theory from a fundamental approach that accounted for all potential interactions in the liquid system. However, this approach requires the consideration of all potential energy interactions, including those that are strongly anharmonic and system- dependent. Without explicit knowledge of the interatomic potential or a small or weak parameter to exploit, the first-principle description of liquids results in a system of coupled non-linear oscillators for which finding solution is impossible with current analytical methods [1]. For this reason, in their canonical treatment of Statistical Mechanics, Landau and Lifshitz assert that it is impossible to derive any general equation describing liquid properties [4]. A major breakthrough was achieved by Stillinger and Weber in the early 1980's [5,6] where, using molecular dynamic simulation (MDS), they introduced and characterized a multi-minimum potential energy surface (PES) of liquid systems as shown schematically in Fig. 1. They found that the PES of a liquid system is rough, *i.e.*, contains many shallow local minima.

A rough PES is visualized in one dimension in Figure 1a. We note that, in contrast to previous works, figure 1a is intended to be a single particle potential. The corresponding dynamics associated with a rough PES include (1) lattice vibrations, which are solid-like, except they generally exhibit anharmonicity due to large displacements from meta-stable equilibrium: (2) large scale diffusion ($\alpha$ transition), which is gas like and describes the hopping motion of the atom from one lattice cage to another as described by Frenkel [1]: (3) small scale diffusion ($\beta$ transitions) corresponding to movement within a local "neighbor cage" and without significant or lasting change to neighbor atoms.

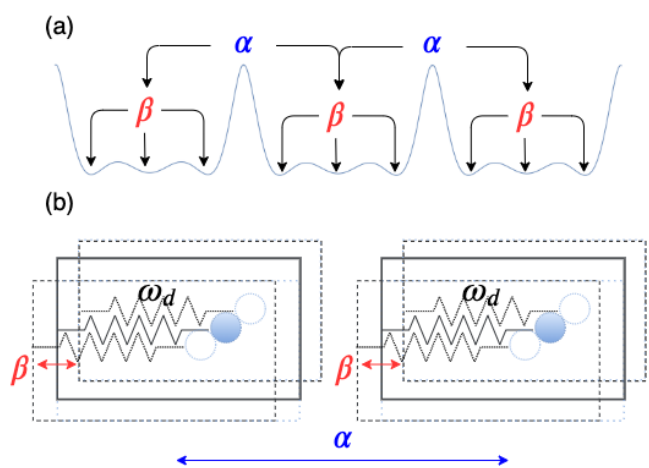

FIG 1: (a) 1D sketch of a liquid's multi-well potential energy surface composed of α and β minima, and (b) a depiction of the PES's corresponding dynamics. A particle vibrates in its lattice cage and diffuses across α and β minima. β transitions represent small scale diffusion corresponding to a changing center of oscillation, and α transitions represent large scale diffusion corresponding to site hopping.

The minima provide a mapping of the PES, which Stillinger and Weber used to factor the partition function into thermal and configurational components [5,6]. Wallace adopted Stillinger's partition function and used entropy of fusion data to quantify the number of inherent structures (*i.e.*, the collection of atomic arrangements at local minima in configuration space) near melting, enabling the evaluation of Stillinger's partition function [7,8]. The inherent structures considered by Wallace involve the $\beta$ minima depicted in Figure 1a. However, for calculating temperature-dependent properties, Wallace ignored large scale diffusion shown by $\alpha$ in Fig. 1.

Trachenko and Brazhkin [1,9–11] avoided the difficult determination of inherent structures by re-introducing J.

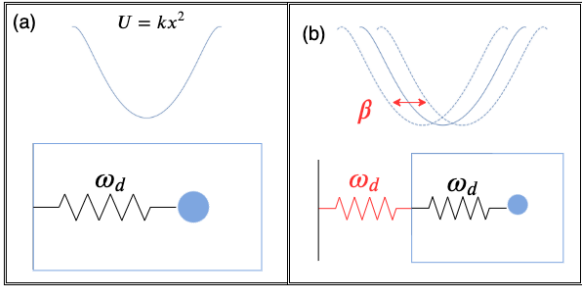

FIG 2: Reconstruction of the PES in Figure 1a by transforming the dynamics. (a) 1D, single-well, single-minima PES corresponding to harmonic vibration within the lattice cage. (b) $\beta$ transitions are represented by springs with frequency $\omega_d$, which are in series with the lattice cage and act to change the center of oscillation resulting in multiple potential minima.

Frenkel's picture of liquid dynamics [12]. A simplified version of this picture describes the microscopic view of atoms as vibrating around equilibrium points at short time scales and hopping to neighboring equilibrium points at larger time scales. Thus, they accounted for large scale diffusion ($\alpha$), but ignored small scale diffusion ($\beta$). Using large scale diffusion, they accurately predicted the decrease in heat capacity at constant volume ($C_v$) as a function of temperature ($T$) by accounting for the loss of transverse phonons with frequencies less than the Frenkel frequency ($\omega_F$). However, their theory was not able to predict the absolute entropy of liquids because they ignored $\beta$ diffusion..

Based on models available in the literature there is a need to develop a theory that accounts for both $\alpha$ and $\beta$ diffusion, and that also predicts the entropy of melting rather than determining it from experimental data. In this letter, we use Stillinger's PES concept to propose a model that transforms the particle dynamics into a form that can be readily treated with statistical mechanics and use Trachenko and Brazhkin's phonon theory to describe the temperature dependence. We then show how this model can offer a simple physical explanation for Richard's melting rule [13] and a material dependent correction to Trouton's rule [14].

In all systems, the thermodynamic and dynamic properties are determined by the nature of the PES [15]. The PES can be characterized by its minima, which correspond to locally stable configurations, and by transition regions connecting those minima. As mentioned, molecular dynamics simulations of monatomic liquids have revealed two distinct types of minima in the PES, one of which is connected by potential barriers much less than $k_bT$ ($\beta$) [6,16], and the other with potential barriers greater than $k_bT$ ($\alpha$), as depicted in Figure 1. The Hamiltonian ($H$) describing the potential energy of a particle in this system can be defined as $= E_{vib} + E_{diff,\beta} + E_{diff,\alpha}$, referring to the vibrational energy, and the small and large scale diffusion, respectively. It is difficult to evaluate because the coupled random walk-like motion inherent in the diffusion terms cannot easily be included in the partition function, so their contribution to phase space and therefore the entropy is unclear in this formulation [1].

In addition, the vibrational and diffusive dynamics are strongly coupled, making the direct determination of an appropriate Hamiltonian and construction of its corresponding partition function a formidable challenge. In this work, we instead transform each particle's motion and its corresponding particle-*PES* into that of a simple harmonic oscillator.

We begin by examining the solid-like local vibrations of a particle within its lattice cage. In general, liquid particles vibrate in a potential well described by both harmonic and higher order (anharmonic) terms that act to soften the spring constant at larger displacements. The anharmonicity associated with lattice vibrations in liquids is not well understood, but it has been shown that it can be neglected for calculating the total entropy near melting [17], so for now we will assume the particle vibrates in a harmonic potential as shown in figure 2a. Anharmonicity is later included as a correction factor for T > $T_m$ where $T_m$ is the melting temperature.

Next, we consider the small-scale $\beta$ diffusion, or the hopping of small energy barriers (< $k_BT$) corresponding to the particles' changing center of oscillation. Small-scale $\beta$ diffusion does not significantly alter the character of the system configuration, so that the change in a particle's neighbor list due to $\beta$ diffusion is difficult to distinguish from that caused by solid-like vibration. Indeed, Rabani et al. [16] were unable to distinguish between the $\beta$ diffusion and solid-like vibrations when observing the decay of neighbor list correlation functions. Provided that the $\beta$ diffusion energy barriers are small, they thus lumped these mechanisms together, both being local perturbations occurring within the domain of a particular particle's local minima. Furthermore, MDS has shown rapid re-crossing of $\beta$ barriers on time scales associated with lattice vibrations, such that a particle's neighbor list correlation function returns to its initial state after $t \approx \frac{2\pi}{\omega_D}$ [16]. In other words, both the vibrational and diffusive motions are responsible for rapid fluctuations in a particle's surroundings, but after $\approx \frac{2\pi}{\omega_D}$ seconds, the particle's local environment has not changed. For this reason, we can consider the particle's motion by imposing a periodic motion on the particles center of oscillation, with angular frequency $\omega_{D,T_m}$, or the Debye frequency evaluated at the melting temperature. Thus, we argue that the dynamics of the particle near melting can be simplified by modeling the small-scale diffusive translational motion as a harmonic spring in series with – and thus altering the center of motion of -- the harmonic lattice-like vibration, as depicted in figure 2b.

Finally, we address the large-scale ($\alpha$) diffusion term in the Hamiltonian describing the hopping motion of the atom from one lattice cage to another, resulting in a large and lasting change in the atoms neighbor list. The hopping rate is described by the Maxwell relaxation time [1] (also known as the Frenkel frequency), $\omega_F(T) = \frac{G}{\eta(T)}$, where $G$ is the high strain rate shear modulus and $\eta(T)$ is the temperature-dependent shear viscosity [9]. At $T_m$, the viscosity of most simple liquids is very high, so the hopping frequency is small and has been shown to be on the order of $\frac{\omega_d}{10}$ or less [1,9,18]. Thus, at $T_m$, the dynamics are dominated by small-scale $\beta$ diffusion and lattice-like vibrations.

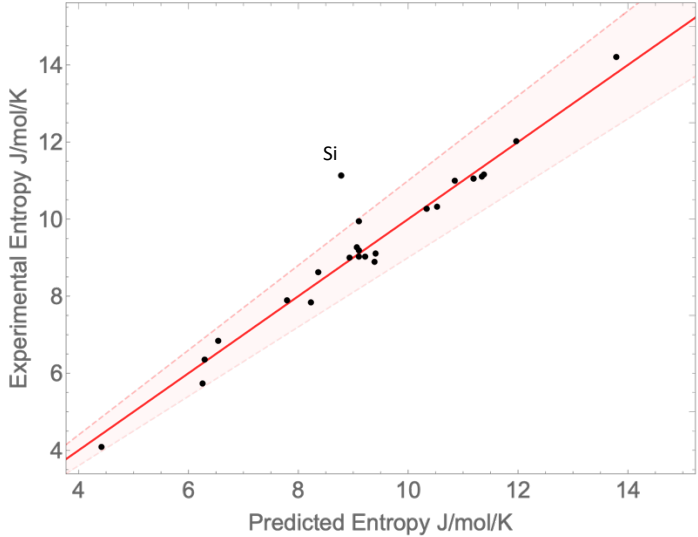

FIG 3: Predicted vs experimental liquid entropy at melt. The solid red line is the $45^o$ line, and the dashed lines represent 10% error. In order of increasing predicted entropy, the dots represent Li, Ar, Ga, Kr, Na, Al, Si, Hg, Xe, Mg, Cd, Zn, In, Sn, K, Cu, Rb, Tl, Ag, Cs, Pb, Au, and W.

For this reason, we neglect $\alpha$ diffusion in our model of liquids dynamics at melt.

In summary, we have modeled each particle in our liquid system as a linear spring representing small-scale diffusion in series with a linear spring describing lattice vibrations (figure 2b) at the melting temperature. These can be combined under an effective spring constant $k_{eff} = \frac{k_{vib} k_\beta}{k_{vib} + k_\beta}$ where $k = m\omega^2$, and re-arranged to get an effective vibrational frequency:

$$\omega_{eff} = \sqrt{\frac{\omega_{vib}^2 * \omega_\beta^2}{\omega_{vib}^2 + \omega_\beta^2}} = \frac{\omega_{D,T_m}}{\sqrt{2}} \quad (1)$$

assuming $\omega_\beta = \omega_{D,T_m} = \omega_{vib}$ as discussed previously. The particle dynamics at $T_m$ can be treated as a simple harmonic oscillator with effective frequency $\frac{\omega_{D,T_m}}{\sqrt{2}}$, as shown in figure 2c. It is important to note that $\omega_{D,T_m}$ refers to the Debye frequency at melting temperature, not at zero temperature. Using the Debye model, we can calculate the entropy associated with the effective Debye frequency at melt as

$S_{T_m} = 4RD\left(\frac{\theta_{D,eff}}{T_m}\right) - 3Rln\left(1 - e^{\frac{\theta_{D,eff}}{T_m}}\right)$, where $\theta_{D,eff}$ the effective Debye temperature and D is is the Debye function. Because we are modeling the liquid state, we can use the high temperature Debye expansion to get the entropy of the liquid at melt:

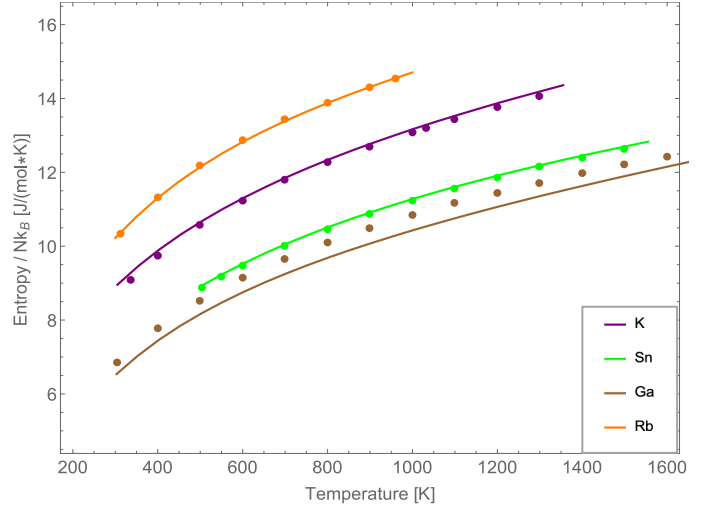

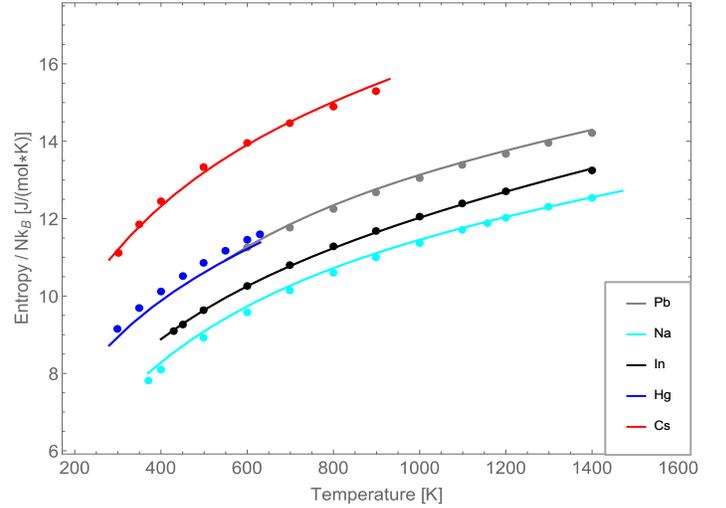

FIG 4: Theory and experimental entropies of various monoatomic liquids versus temperature. Experimental entropy values were taken from Selected Values of the Thermodynamic Properties of the Elements at a pressure of 1 atm. Solid lines represent theory, whereas dots represent experimental data. Input parameters and their sources are provided in SI section 4.

$$S_{T_m} = 4R + 3Rln\left(\frac{k_b T_m}{\hbar \frac{\omega_{D,T_m}}{\sqrt{2}}}\right) \quad (2)$$

We emphasize that this expression is in agreement with the classical result for entropy derived from the single-oscillator canonical partition function, given the relation between the Einstein and Debye frequency of the classical oscillator, $\omega_E = e^{-\frac{1}{3}}\omega_D$ [17].

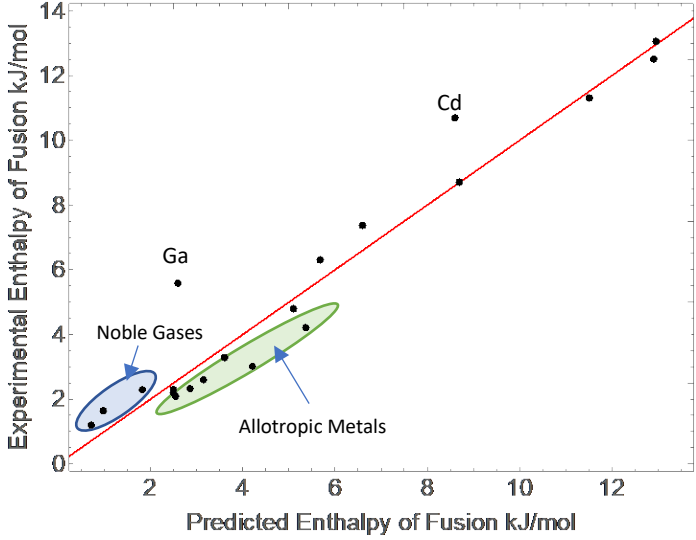

FIG 5: Predicted vs experimental enthalpy of fusion. The solid red line is the $45^o$ line, and the dashed lines represent 10% error. The blue oval contains the noble gasses, and the green oval contains the allotropic metals. In order of increasing predicted enthalpy of fusion, the dots represent Xe, Ar, Kr, K, Ga, Pb, In, Cs, Hg, Na, Rb, Cu, Li, Ag, Mg, Zn, Al, Tl, Cd, Au. We note that Si, which has a predicted and experimental $\Delta H$ of 15 and 39 kJ/mol, respectively is not shown

A comparison of the entropy at melt (equation 2) to experimental data from Selected Values of the Thermodynamic Properties of the Elements [19] for 24 monatomic liquids is plotted in Figure 3 Debye temperatures evaluated at the crystal melt and density were taken from [17], and we observe that with just this single input property equation 2 predicts entropy at melt to within 10% of experimental values for 23 of the 24 liquids. The notable outlier is silicon. Wallace identifies silicon as an "anomalous melting" element because it is shown to undergo significant change in electronic structure from crystal to liquid [7,17], which equation 2 does not account for.

To accurately predict the thermodynamic properties at $T > T_m$ under constant pressure, we must include the entropy increase due to expansion and anharmonicity, and the entropy decrease associated with the loss of transverse phonons, which account for the effect of large-scale diffusion. This can be written as $S(T) = S_{T_m} + \int_{T_m}^{T} \frac{C_p}{T} dT$, where $C_p = 3R + C_{exp} + C_{anharmonic} - C_{loss}$.[20] Expansion heat capacity is expressed as $C_{exp} = MB\alpha_V^2 T$ where $M$ is the molar volume, $B$ is the fluid's bulk modulus and $\alpha_V$ is the fluid's volumetric thermal expansion coefficient. The anharmonic heat capacity can be approximated as $C_{anharmonic} = 3R\alpha_V T$.[10,21] The heat capacity associated with the large-scale diffusion, or the loss of transverse phonons is $C_{loss} = \frac{d}{dT}\left[RT\left(\frac{\omega_F(T)}{\omega_D}\right)^3\right]$.[10] Therefore $S(T)$ can then be written as:

$$S(T) = S_{T_m} + 3R\ln\left(\frac{T}{T_m}\right) + MB\alpha_V^2(T - T_m)$$
$$+3R\alpha_V(T - T_m) - R\int_{T_m}^{T}\left\{\frac{1}{T}\frac{d}{dT}\left[RT\left(\frac{\omega_F(T)}{\omega_D}\right)^3\right]\right\}dT \quad (3)$$

The phonon loss term (final term in equation 3) can be approximated as $S_{Loss}(T) = -R\left(\frac{\omega_F(T)}{\omega_D}\right)^3$ with less than 2% total error on S(T) (see supplementary information, (SI) section 1). All other thermodynamic properties of interest can be determined using appropriate thermodynamic relations with (3). For $S(T)$ at constant volume, the expansion term (3rd term) in Eqn. 3 should be neglected. Details on calculation of $\omega_F$, including differences from values used by Trachenko and Brazhkin [9], are provided in the SI. Comparison of equation 3 to experimental entropy data as a function of temperature is plotted in Figure 4. Besides gallium, which also has a significant change in electronic structure like silicon [17], there is excellent agreement between experiment and model. Constant pressure and constant volume heat capacities evaluated as $C_{P,V} = T\left(\frac{\partial S}{\partial T}\right)_{V,P}$ are compared to experimental data in SI section 2 based on our calculation of $\omega_F$.

In addition, equation 2 can be combined with the Debye model for crystals to calculate the entropy and enthalpy of fusion:

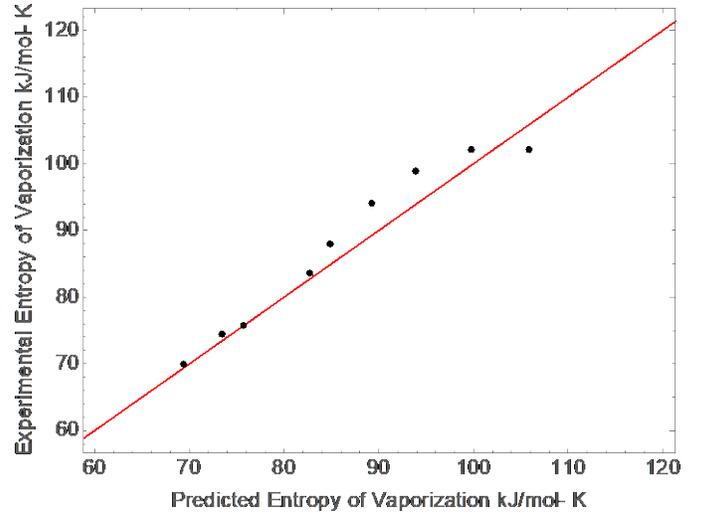

FIG. 6: Predicted vs experimental entropy of vaporization. The solid red line is the $45^o$ line, and the dashed lines indicate 10% error. The black dashed horizontal line shows Trouton's Rule. In order of increasing predicted entropy, the dots represent Ga, Sn, Pb, In, Cs, Rb, Hg, Na, and K.

$$\Delta S_M = 3R\ln\left(\frac{\sqrt{2}\omega_{D,T_m}^S}{\omega_{D,T_m}^L}\right) \quad (4)$$

where $\omega_{D,T_m}^S$ is the Debye frequency in the solid state and $\omega_{D,T_m}^L$ is the Debye frequency in the liquid state at melt. Thus far, we have assumed (and it has been shown [22]) that for metals $\omega_{D,T_m}^L \approx \omega_{D,T_m}^S$. In this case, equation 4 reduces to $\Delta S \approx 1.1R$ which is the empirical value used in Richard's Melting Rule [13]. Thus, we offer the diffusion-vibration oscillator dynamics near melt as a possible explanation for this empirical observation.

In Figure 5, we compare equation 4 to the enthalpy of fusion of 20 different monatomic liquids given by $\Delta H_M = T_M \Delta S_M$, assuming that $\omega_D^S = \omega_D^L$. Equation 4 systematically overpredicts for allotropic metals, shown in the green oval. The allotropic metals undergo solid-solid phase transitions before melting which increase the enthalpy of the solid phase at melt [13]. Equation 4 also systematically underpredicts for the noble gases, shown in the blue oval. This is expected from our formulation because it has been shown [22] that $\omega_{D,T_m}^L \neq \omega_{D,T_m}^S$ for the noble gasses. Using $\omega_{D,T_m}^S$ as a substitute for $\omega_{D,T_m}^L$, which is unknown for most liquids, is inaccurate in such instances. Furthermore, the viscosity of the noble gasses is low at melt such that $\omega_{F,T_m} \approx \omega_{D,T_m}$. Therefore, large-scale diffusive dynamics become important, which our model neglects at melt.

Finally, we compare our model with experimental data for the entropy of vaporization ($\Delta S_V$), which is typically given by the well-known Trouton' rule [14] that states $\Delta S_V \approx 88$ J/mol –K. We used experimental entropy data for the gas phase [19] and subtracted it from equation 3 evaluated at the boiling temperature to predict $\Delta S_V$, as plotted in Fig. 5. Examination of Fig. 5 reveals that our model predicts $\Delta S_V$ very well (mean absolute error of 2.42 J/mol-K) whereas Trouton' rule, which is independent of material properties, gives a constant value. Thus, we have shown that equation 3 gives accurate thermodynamic values over the entire liquid range at atmospheric pressure.

In summary we have developed a simple analytical model to predict thermodynamic properties of monatomic liquids by accounting for both small scale and large-scale diffusion. In our calculations we assumed $\omega_D^S \approx \omega_D^L$ which is valid for most metals but leads to bad predictions when there is a strong deviation in the frequency from the solid to the liquid state (e.g. noble elements). Data for $\omega_D^L$ is rare, and future efforts in determining $\omega_D^L$ are needed to more accurately test the model for such liquids. Moreover, we restricted our analysis to monatomic systems to isolate the thermodynamic contributions from the inter-molecular interactions, and therefore make a simpler and more meaningful comparison to our model. Further work must be done to include the effects of intra-molecular interactions in order to test the model for multiatomic systems.

Authors acknowledge the support of the Laboratory Directed Research and Development Program (LDRD) at Lawrence Berkeley National Laboratory under contract # DE-AC02-05CH11231